\documentclass[reqno,tbtags,a4paper,11pt]{amsart}

\usepackage[utf8]{inputenc}
\usepackage{amsmath,amssymb,amscd,amsthm,a4wide}
\usepackage{graphicx} %make graphicx
\usepackage[all]{xy}
\usepackage{color}
\usepackage{xcolor}

\newtheorem{thm}{Theorem}%[section]

\newtheorem{lemma}[thm]{Lemma}

\theoremstyle{definition}

\theoremstyle{remark}

\newtheorem*{conj}{\bf Conjecture}

\newcommand{\cA}{\mathcal A}

\newcommand{\cI}{\mathcal I}
\newcommand{\cJ}{\mathcal J}

\newcommand{\cS}{\mathcal S}

\newcommand{\bbC}{\mathbb C}

\newcommand{\bbN}{\mathbb N}

\newcommand{\bbZ}{\mathbb Z}

\title{Integrable Volterra hierarchies over nonabelian algebras } 
\author{J.~P.~Wang$^\#$, S.~Carpentier$^\dagger$ and A.~V.~Mikhailov$^\natural$}
\date{}
\address{$^\#$   Ningbo University, Ningbo 315211,
People's Republic of China
 }
\email{wangjingping1@nbu.edu.cn }
\address{$^\dagger$   Seoul National University, South Korea
 }
\email{sylcar@snu.ac.kr }
\address{$^\natural$ University of Leeds,\, UK}
\email{a.v.mikhailov@leeds.ac.uk}
\thanks{This paper is a translation of a Russian manuscript accepted for publication in Russian Mathematical Surveys.\\ \indent SC is supported by BK21 Seoul National University Mathematical Sciences Division.
JPW is supported by NSFC grant No. 12571265 and the Ningbo University Research Start-up Fund.}

\begin{document}
\maketitle

Known integrable systems with noncommutative dependent variables are typically formulated over free associative algebras, quantum algebras, or Grassmann algebras. For differential-difference integrable equations, we identify a new class of noncommutative algebras that is compatible with the dynamics and can be positioned between quantum and free algebras.
In this brief communication, we consider reductions of the nonabelian Volterra hierarchy \cite{bog91} to new algebras $\cA_\cJ$ and $\cA_{\hat{\cJ}}$. This approach extends to a broad class of integrable systems, including the Toda lattice, the Ablowitz-Ladik system, and many others.

Let $\cA=(\bbC,\, \{u_n\}_{n\in\bbZ},\, \cS)$ be a unital free difference algebra, generated by the infinite set of noncommuting variables, and equipped with the shift automorphism $\cS$ defined by  $\cS(u_n)= u_{n+1}$. The involution $^+$, anti-automorphism of $\cA$ (analogue of Hermitian conjugation),    is defined by   $ u_n^+=u_n,\ (ab)^+=b^+a^+$ and $(\alpha)^+=\bar{\alpha}$ for any $u_n, a,b\in\cA$ and $\alpha\in\bbC$.

Equations of the Volterra hierarchy
\begin{equation}\label{vol0}
\partial_{t_1} u_n = u_{n+1}u_n - u_n u_{n-1}, \quad\partial_{t_\ell}u_n=K^{(\ell)}(u_{n-\ell},\ldots ,u_{n+\ell}),\ \ell\in\bbN,\   n \in \mathbb{Z},
\end{equation}
are shift invariant, and define commuting derivations $\partial_{t_\ell}$ of $\cA$. They are fully determined by the $n=0$ component $\partial_{t_\ell}u=K^{(\ell)}$. Here and in the following $u=u_0$. The \emph{adjoint} integrable hierarchy $\partial_{\tau_\ell}u= (K^{(\ell)})^+$ is not compatible with (\ref{vol0}) since $[\partial_{t_\ell},\partial_{\tau_m}]\ne 0$ in $\cA$.

{\bf 1.} {\it Reduction to the algebra $\cA_\cJ$.} We denote by $\mathcal{I}$ the two-sided difference ideal of $\cA$ generated by commutators
$$\cI=\langle u_n u_m-u_m u_n\,;\, |n-m|\ne 1,\, n,m\in\bbZ\rangle,$$
and define a larger ideal $\cJ\supset \cI$ by adding the permutation relations
\begin{eqnarray*}
   \cJ\,&=&\cI+\langle  u_{n+1}u_nu_{n+2}-u_{n+2}u_nu_{n+1}\,;\,n\in\bbZ\,\rangle  .
\end{eqnarray*}
These ideals are self-adjoint.
The ideal $\cI $ is not $\partial_{t_\ell}$--stable, i.e.  $\partial_{t_\ell}(\cI ) \not\subset \cI $,  hence the Volterra hierarchy can not be reduced to the quotient algebra $\cA_\cI=\cA/\cI$.
In contrast, the ideal  $\cJ$ is remarkably $\partial_{t_1}$--stable, so the Volterra system is well defined on $\cA_\cJ=\cA/\cJ$.
\begin{thm}  {\rm (1)}  $\cJ$ is the minimal $\partial_{t_1}$--stable ideal containing $\cI$.  {\rm (2)}    The ideal $\cJ$ is a minimal extension of $\cI$ that implies the compatibility of the systems
\begin{equation}\label{vol1}
   \partial_{t_1} u=u_{1}u-uu_{-1},\qquad \partial_{\tau_1} u=uu_{1}- u_{-1}u. 
\end{equation}
\end{thm}
Using the Lax representation for the Volterra hierarchy on a free algebra \cite{bog91}, we proved:
\begin{lemma}
The formal series
\begin{equation*}
    T(\lambda)=1+\sum_{m\ge1}\lambda^m T^{(m)},\quad
T^{(m)}=\sum_{(k_1,\ldots,k_m)\in\Gamma_m} u_{k_1}\cdots u_{k_m},\quad \Gamma_m=\{(k_1,\ldots,k_m)\in\mathbb{Z}^m\, ;\, k_j>k_{j+1}+1\ \forall j\},
\end{equation*}
satisfies \(\partial_{t_\ell}T(\lambda)=0\) and thus $\partial_{t_\ell}T^{(m)}=0$.
%generates the nonlocal first integrals \(T^{(m)},\ \partial_{t_\ell}T^{(m)}=0\) of the Volterra hierarchy (\ref{vol0}) on \(\mathcal{A}\).
\end{lemma}
In the algebra $\cA_\cJ$ it leads to much stronger results.
\begin{thm}\label{th3} Modulo $\cJ$: {\rm (1.)} \(T(\lambda)T(\mu)=T(\mu)T(\lambda)\), hence \([T^{(n)},T^{(m)}]=0\). {\rm (2.)} The expansion \( \log T(-\lambda) =-\sum_{\ell=1}^\infty \frac{\lambda^\ell}{\ell} H_\ell \) defines first integrals \( H_\ell=\sum_{k\in\bbZ}\cS^k(h_\ell) \) of (\ref{vol1}) with $\cJ$--local densities $h_\ell\in\cA$.  {\rm (3.)} \([H_\ell,H_m]=0\). \quad {\rm (4.)} \(H_\ell=H_\ell^+\). \quad {\rm (5.)} \([H_1,u]=K^{(1)}-K^{(1)+}\). \end{thm}
The $\cJ$--locality of $h_\ell$ means that 
$\forall a \in \mathcal{A},\ \exists N_a\in\bbN$ such that $ [h_\ell, \mathcal{S}^k(a)] \in \mathcal{J}$   for $ |k|>N_a$.

{\bf 2.} {\it Reduction to the algebra $\cA_{\hat{\cJ}}$.}  For the first system of the even Volterra sub-hierarchy $\partial_{t_{2\ell}}u=K^{(2\ell)}$ 
\begin{equation}\label{vol2}
    \partial_{t_2}u=K^{(2)}=u_2u_1u+u_1^2u+u_1u^2-u^2u_{-1}-uu_{-1}^2-uu_{-1}u_{-2},
\end{equation}
we consider two-sided ideals of $\cA$:
\begin{eqnarray*}
\hat\cI&=&\langle u_n u_m+u_m u_n\,;\, |n-m|\ne 1,\, n,m\in\bbZ\rangle,\\
\hat\cJ\,&=&\hat{\cI}+\langle  u_{n+1}u_nu_{n+2}-u_{n+2}u_nu_{n+1}\,;\,n\in\bbZ\,\rangle .
\end{eqnarray*}
 \begin{thm}\label{th4}   $\hat{\mathcal{J}}$ is the minimal $\partial_{t_2}$--stable ideal containing $\hat{\mathcal{I}}$. Modulo $\hat{\mathcal{J}}$:
 {\rm (1.)}      $[\partial_{t_2},\partial_{\tau_2}]= 0$. {\rm (2.)} For
 $ Q(\lambda)=T(\lambda)T^+(-\lambda)$,  $Q(\lambda)Q(\mu)=Q(\mu)Q(\lambda)$.
 {\rm (3.)} The expansion
 $\log Q(\lambda)=-\sum_{\ell=1}^\infty \frac{1}{\ell}\lambda ^{2\ell} \hat{H}_{2\ell}$ defines $\hat{\cJ}$--local first integrals $\hat{H}_{2\ell}$ for (\ref{vol2}). {\rm (4.)} $[\hat{H}_{2\ell},\hat{H}_{2m}]=0$. {\rm (5.)} $\hat{H}_{2\ell}=\hat{H}_{2\ell}^+$.  {\rm (6.)}   $[\hat{H}_2,u]=K^{(2)}- K^{(2) +}$.
\end{thm}
Thus, the system (\ref{vol2})
is well defined on the algebra $\cA/\hat\cJ$, admits the infinite set of local self-adjoint first integrals ($\partial_{t_2}(H_{2\ell})=0$) and is compatible with the adjoin equation $\partial_{\tau_2}u=K^{(2)+}$ modulo the ideal.

\iffalse
 We conjecture $\partial_{t_\ell}$--stability of $\cJ$,  $[\partial_{t_\ell}, \partial_{\tau_m}]=0$ and $[H_\ell,u]=K^{(\ell)}- K^{(\ell) +}$   in $\cA_\cJ$ (respectively  $\partial_{t_{2\ell}}$--stability of $\hat \cJ$,  $[\partial_{t_{2 \ell}}, \partial_{\tau_{2m}}]=0$ and $[H_{2\ell},u]=K^{(2\ell)}- K^{(2\ell)+}$  in $\cA_{\hat\cJ}$) for all $\ell, m\in\bbN$.
\fi

{\bf 3.} {\it Connection with quantum algebras.} It was shown in \cite{CMW22} that: {\rm (i.)} The ideal $\cI_{\omega}\subset \cA$:
\[
\cI_\omega=\cI+\langle u_{n}u_{n+1}-\omega u_{n+1}u_n\,;\,n\in\bbZ,\, \bar{\omega}=\omega^{-1} \in \mathbb{C}^*\rangle
\]
is $\partial_{t_\ell}$--stable for all $\ell$. It defines the standard quantisation of the Volterra hierarchy (\ref{vol0})  \cite{Inka}. {\rm (ii.)} The ideal 
\[
\hat{\cI}_\omega=\hat{\cI}+\langle u_{n}u_{n+1}-(-1)^n\omega u_{n+1}u_n\,;\,n\in\bbZ,\, \bar{\omega}=\omega^{-1} \in \mathbb{C}^*\rangle
\]
is $\partial_{t_{2\ell}}$--stable, leading to a non-standard quantisation \cite{Mik20}. 
\begin{lemma}
{\rm (i.)} $\cJ\subset \cap_{\omega\in\bbC^*}\cI_\omega$. {\rm (ii.)} $\hat\cJ\subset \cap_{\omega\in\bbC^*}\hat{\cI}_\omega$. 
\end{lemma}
\begin{thm} For all $\ell,m\in\bbN$:
{\rm (i.)}  In the quantum algebra 
$\cA/\cI_\omega$:
\[  [H_\ell ,H_m]= 0 ,\quad \partial_{t_\ell}H_m=0,\quad [\partial_{t_\ell}, \partial_{\tau_m}]=0,\quad K^{(\ell)+}=\omega^\ell K^{(\ell)},\quad (1-\omega^\ell)\partial_{t_\ell}u=[H_\ell,u].\]
{\rm (ii.)} In the quantum algebra  $\cA/\hat{\cI}_\omega$:
\[[\hat{H}_{2\ell},\hat{H}_{2m}]=0,\quad \partial_{t_{2\ell}}\hat{H}_{2m}=0,\quad [\partial_{t_{2 \ell}}, \partial_{\tau_{2m}}]=0,\quad K^{(2\ell)+}=\omega^{2\ell} K^{(2\ell)},\quad (1-\omega^{2\ell})\partial_{t_{2\ell}}u=[\hat{H}_{2\ell},u].\]
\end{thm}
The explicit forms of the quantum Hamiltonians for both the standard and non-standard quantisations of the Volterra hierarchy were presented in \cite{CMW22}. Theorems~\ref{th3} and~\ref{th4} provide an alternative approach to the computation of these Hamiltonians. For instance, it follows from Theorem~\ref{th3} that %in the quantum algebra $\cA/\cI_\omega$ 
\[h_1=u,\ h_2=u^2+uu_1+u_1u,\  h_3=u^3+u^2 u_1+uu_1u+u_1u^2+uu_1^2+u_1uu_1+u_1^2 u+uu_1u_2+u_1u_2u+u_2u_1u,
\]
which yields, in the quantum algebra $\cA/\cI_\omega$, the three Hamiltonians explicitly obtained in \cite{Inka}:
\[
H_1=\sum_{n\in\bbZ}u_n,\quad H_2=\sum_{n\in\bbZ} u_n^2+(1+\omega)u_{n+1}u_n,\quad 
H_3=\sum_{n\in\bbZ}u_n^3+(1+\omega+\omega^2)(u_{n+1}^2u_n+u_{n+1}u_n^2+u_{n+2}u_{n+1}u_n).
\]
\begin{conj}
For all $\ell, m\in\bbN$: {\rm (i.)}  $\cJ$ is $\partial_{t_\ell}$--stable, and in $\cA_\cJ$ one has $[\partial_{t_\ell}, \partial_{\tau_m}]=0$ and $[H_\ell,u]=K^{(\ell)}- K^{(\ell) +}$;
{\rm (ii.)}  $\hat \cJ$ is $\partial_{t_{2\ell}}$--stable,  and in $\cA_{\hat\cJ}$ one has $[\partial_{t_{2 \ell}}, \partial_{\tau_{2m}}]=0$ and $[\hat{H}_{2\ell},u]=K^{(2\ell)}- K^{(2\ell)+}$. 
\end{conj}
This conjecture has been verified for $\ell,m\le 4$ and holds in the quantum cases for all $\ell, m\in\bbN$.
\iffalse
and therefore in the quantum algebra $\cA/\cI_\omega$ 
\[
H_1=\sum_{n\in\bbZ}u_n,\quad H_2=\sum_{n\in\bbZ} u_n^2+(1+\omega)u_{n+1}u_n,\quad 
H_3=\sum_{n\in\bbZ}u_n^3+(1+\omega+\omega^2)(u_{n+1}^2u_n+u_{n+1}u_n^2+u_{n+2}u_{n+1}u_n). 
\]

It follows from Theorem~\ref{th4} that in  
$\cA/\hat{\cI}_\omega$: $\hat{H}_2=\sum_{n\in\bbZ} u_n^2+(1+(-1)^n\omega)u_{n+1}u_n$ and
\begin{eqnarray*}
    \hat{H}_4&=&\sum_{n\in\bbZ}\Big( u_n^4+(1-\omega^4)u_{n+1}u_n^2u_{n-1}+(1+\omega^2)(1+(-1)^n\omega+\omega^2)u_{n+1}^2u_n^2\Big)\\
    &+&(1+\omega^2)\sum_{n\in\bbZ}(1+(-1)^n\omega)(u_{n+1}^3u_n+
u_{n+1}u_n^3+u_{n+1}u_nu_{n-1}^2+u_{n+2}^2u_{n+1}u_n+u_{n+3}u_{n+2}u_{n+1}u_n) .
\end{eqnarray*}
\fi

%\bibliographystyle{unsrt}
%\bibliography{shortbib}

\end{document}